\documentclass{pasj00}

\begin{document}
\SetRunningHead{Nagayama et al.}{Water Masers in Onsala 1}
\Received{2007/09/04}
\Accepted{2007/11/08}

\title{VLBI Observations of Water Masers in Onsala 1:\\
       Massive Binary Star Forming Site?}

\author{Takumi \textsc{Nagayama},\altaffilmark{1}
        Akiharu \textsc{Nakagawa},\altaffilmark{2}
        Hiroshi \textsc{Imai},\altaffilmark{2}
        Toshihiro \textsc{Omodaka},\altaffilmark{2}\\
        and Yoshiaki \textsc{Sofue}\altaffilmark{2}}
\altaffiltext{1}{Graduate School of Science and Engineering, 
                 Kagoshima University, \\
                 1-21-35 K\^orimoto, Kagoshima 890-0065}
\altaffiltext{2}{Faculty of Science, Kagoshima University, 
                 1-21-35 K\^orimoto, Kagoshima 890-0065}
\email{nagayama@astro.sci.kagoshima-u.ac.jp}


\maketitle


\begin{abstract}
We present proper motions of water masers toward 
the Onsala 1 star forming region, 
observed with the Japanese VLBI network 
at three epochs spanning 290 days.
We found that there are two water maser clusters (WMC1 and WMC2)
separated from each other by \timeform{1".6} 
(2900 AU at a distance of 1.8 kpc).
The proper motion measurement reveals that
WMC1 is associated with a bipolar outflow 
elongated in the east-west direction with 
an expansion velocity of $69 \pm 11$ km s$^{-1}$.
WMC1 and WMC2
are associated with two 345 GHz continuum dust emission sources,
and located \timeform{2"} (3600 AU) east from 
the core of an ultracompact H\emissiontype{II} region
traced by 8.4 GHz radio continuum emission.
This indicates that
star formation activity of 
Onsala 1 could move from the west side
of ultracompact H\emissiontype{II} region to 
the east side of two young stellar objects associated with the water masers.
We also find that 
WMC1 and UC H\emissiontype{II} region 
could be
gravitationally bound.
Their relative velocity along
the line of sight 
is $\sim 3$ km s$^{-1}$, and total mass is 
$\sim 37 \MO$.
Onsala 1 seems to harbor a binary star 
at different evolutionary stage.
\end{abstract}


\section{Introduction}

Star forming regions are often associated with water maser emissions,
and the Very Long Baseline Interferometry (VLBI) monitoring observations 
of the water maser provide a unique tool 
to study the structure and kinematics of star forming region. 
Analyses of spatial positions, Doppler velocities, and proper motions
of water masers have revealed the 3-D gas kinematics often in the
vicinity of young stellar objects (YSOs)
(e.g., Orion-KL: \cite{gen81}; Cepheus A: \cite{tor01}; 
G192.16$-$3.84: \cite{imai06}; G24.78+0.08: \cite{mos07}).

Onsala 1 star forming region (hereafter ON1) has
an ultracompact (UC) H\emissiontype{II} region observed in
radio continuum emissions at 8.4 GHz and 23.7 GHz (\cite{zheng}; \cite{argon}).
Centimeter radio continuum luminosity
of this UC H\emissiontype{II} region
is $10^{4.2} \LO$, 
and indicates an exciting star of ZAMS type B0 (\cite{macleod}).
\citet{kumar2004} reported the presence of multiple outflows from
the UC H\emissiontype{II} region.
One outflow traced by $^{12}$CO $J$ = 2--1 line
is elongated in
the east-west direction
with a velocity of 12 km s$^{-1}$
(respect to the systemic velocity of ON1 $V_{\rm LSR} =$ 12 km s$^{-1}$),
and
other possible outflow 
traced by H$^{13}$CO$^{+}$ $J$ = 1--0 line is elongated in
the northeast-southwest direction
with a velocity of 4.5 km s$^{-1}$ (\cite{kumar2004}).
The H$^{13}$CO$^{+}$ outflow is, 
however on the other hand, interpreted as a rotating disk
by the observations 
in the NH$_3$ and same H$^{13}$CO$^{+}$ lines (\cite{zheng}; \cite{lim02}).
These molecular line observations suggest that
there are several YSOs in or 
around the UC H\emissiontype{II} region of ON1.

Submillimeter continuum emission at 345 GHz (0.85 mm) using
the Submillimeter Array (SMA) have resolved
two components 
separated in the northeast--southwest direction (\cite{su}).
The 345 GHz continuum emissions are 
elongated
in the east side of the UC H\emissiontype{II} region.
It is known that YSOs at the youngest stage often 
show spectral energy distribution
with peak flux at submillimeter wavelength,
originating from the dust envelopes (\cite{and93}).
The strong submillimeter emissions from these two sources would, therefore,
indicate that they are the youngest YSOs contained in ON1.
Thus,
YSOs at different evolutionary stages seem to be found in ON1.

VLBI observation of the water masers 
(\cite{downes}) shows that
ON1 has two clusters of water masers separated by $\sim$\timeform{2"} 
(3600 AU at our assumed distance of 1.8 kpc)
in the northeast-southwest direction.
These two clusters appear to have offsets from
the core of the UC H\emissiontype{II} region traced by
23.7 GHz continuum emission (\cite{zheng}).
This fact suggests that 
the water masers are associated with
the newly YSOs which formed around the UC H\emissiontype{II} region.
In order to confirm this,
new multi-epoch VLBI observations
to measure the proper motions of water masers are necessary.

The VLBI observation of \citet{downes} detected
water masers at $V_{\rm LSR} = 5$--23 km s$^{-1}$.
\citet{kurtz} found that 
the water maser emission of ON1 has a larger velocity range
($V_{\rm LSR} = -53$ to 63 km s$^{-1}$), 
and suggested that these high velocity components would
give an important information about other star formation signposts in ON1.

Here, we present the distribution and proper motions 
of 22.2 GHz water masers towards ON1.
Section 2 describes the observations
with the Japanese VLBI network (JVN, e.g., \cite{doi}; \cite{imai06})
and data reduction.
Section 3 shows the distribution and proper motion of the water masers.
Section 4 
discusses the driving sources of water masers 
and structure of ON1 
through comparison with previous observations.
The distance to ON1 is uncertain, 
however, a near kinematic distance of 1.8 kpc is 
favored by the most authors
(e.g. \cite{macleod}; \cite{kumar2004}).
We therefore adopt the distance of 1.8 kpc to this source.


\section{Observations}

The VLBI observations of ON1 were made on March 24, and 
June 1, in 2005 and on January 8, in 2006 using 
five or six telescopes of the JVN composed of 
four 20-m telescopes of the VLBI Exploration of Radio Astrometry (VERA)
of the National Astronomical Observatory of Japan (NAOJ),
a 45-m telescope of the Nobeyama Radio Observatory (NRO), and
a 34-m telescope of the National institute of Information
and Communications Technology (NiCT) at Kashima.
ON1 and a calibrator source (ICRF J192559.6+210626) 
for clock parameter correction were observed for 
10 hours in total per observing epoch.
Left-hand circular polarization signals were recorded with
the SONY DIR1000 recorder with a data rate of 128 Mbps and
in two base band channels with a band width of 16 MHz each,
covering a radial velocity span of 215.7 km s$^{-1}$.
The data correlation was made with the Mitaka FX correlator (\cite{chikada}).
The correlation outputs consisted of 1024 velocity channels,
yielding frequency and velocity resolutions
of 15.625 kHz and 0.21 km s$^{-1}$, respectively.
 
Data reduction was performed using 
the Astronomical Image Processing System (AIPS)
of National Radio Astronomy Observatory (NRAO).
Calibrations of the clock parameters,
bandpass characteristics, visibility amplitudes, and phase visibility phases as well as
velocity tracking were carried out in a
standard manner.
The clock parameters (clock offset and clock rate offset) were calibrated
using the residual delay and delay rate for a calibrator source
which was observed every hour.
Bandpass calibration was made using auto correlation spectra of the continuum
source.
The amplitude calibration was made using the system noise temperatures;
they were evaluated using the ``R-Sky'' method, by observing a reference black
body at the beginning of each scan (typically every hour).
Observed frequencies of maser lines were converted to 
the local-standard-of-rest (LSR) velocities
using the rest frequency of 22.235080 GHz for H$_2$O $6_{16}-5_{23}$ transition.
For phase calibrations, the visibilities of all velocity channels were
phase-referenced to the reference maser spot 
at a LSR velocity of 16.5 km s$^{-1}$, 
which is one of the brightest spots and
it exhibits no sign of structure according to the closure phase deviation from zero.
A typical size of the synthesized beam was $\sim$ 1 milliarcseconds (mas)
in the three observations (see Table \ref{tab:1}). 

We identified all emission components stronger than 7 times
the rms noise level in the images for each spectral channel.
Identification of a water maser feature, 
which represents a physical feature and 
consisting of a cluster
of maser spots or velocity components 
was made by
the procedure shown in several previous papers (e.g. \cite{imai}).
We identified as the feature 
consisting of spots within a beam size of approximately 1 mas (1.8 AU).
The feature position is defined as a 
brightness peak in the feature.
Position uncertainties for the features were typically 0.05 mas.
We detected 21, 18, and 23 maser features in the three epochs, respectively.


\section{Results}

\subsection{Water Maser Spectrum}

\begin{figure*}[h]
\begin{center}
\FigureFile(80mm,100mm){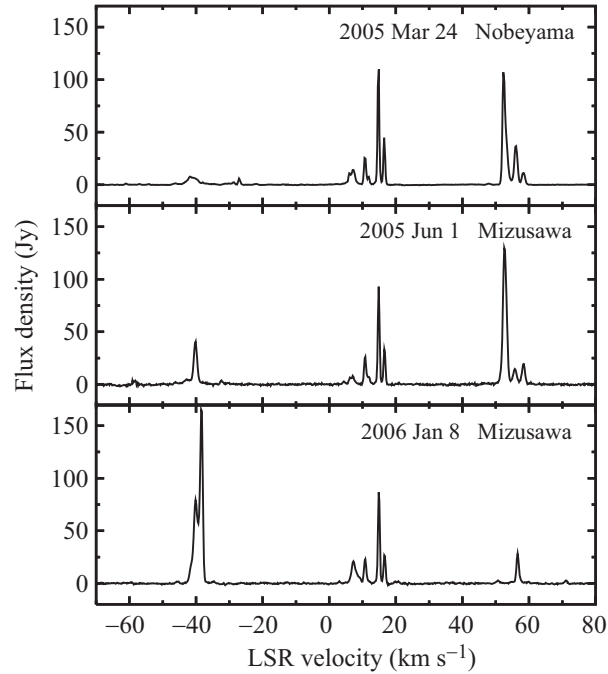}
\end{center}
\caption{The spectra of total flux of ON1 water masers,
         obtained in the three epochs
	 with the Nobeyama 45-m and the Mizusawa 20-m telescopes 
	 The velocity resolution is equal to 
         the channel width of 0.21 km s$^{-1}$.}
\label{fig:1}
\end{figure*}

Figure \ref{fig:1} shows total power spectra
of the ON1 water masers obtained 
with the Nobeyama 45-m and Mizusawa 20-m telescopes.
There are low-velocity components ($V_{\rm LSR} \simeq 4$--22 km s$^{-1}$)
near the systemic velocity of ON1
observed in the NH$_3$ line ($V_{\rm LSR} = 10$--12 km s$^{-1}$; \cite{zheng})
and,
in addition, 
high-velocity blue-shifted 
($V_{\rm LSR} \simeq -60$ to $-20$ km s$^{-1}$),
and red-shifted ($V_{\rm LSR} \simeq 50$--60 km s$^{-1}$) components.
The blue- and red-shifted components
symmetrically appear around the
systemic velocity.
We have detect most of the 
velocity components seen
in the previous observations (\cite{downes}; \cite{kurtz}).

The peak flux density of the low-velocity components 
is typically 100 Jy, and do not change from epoch to epoch.
On the other hand, fluxes of the high-velocity components 
are time-variable.
The blue-shifted components had 8 Jy in March 2005,
and increased to more than 150 Jy in January 2006.
These blue-shifted components were already found about 30 years ago
by a singledish observation (\cite{genzel77}).
However, they were too weak
to be mapped by the VLBI in 1977 (\cite{downes}).
The red-shifted components were strong (approximately 100--150 Jy)
during our first and second VLBI observations (March--June 2005). 
However, they decreased to 30 Jy in January 2006. 
These redshifted components were not detected in 1977 and 1987, 
(\cite{genzel77}; \cite{cesaroni}), 
and detected for the first time in 1995 (\cite{kurtz}).

Statistics of water masers show that
high-velocity components are generally highly variable,
and are as weak as 0.1--10\% of the low-velocity component
(\cite{genzel77}).
Similarly, 
the high-velocity component in ON1 is also highly variable.
However, the high-velocity component is sometimes 
stronger than the low-velocity component.
Its intensity is
10--200\% of the low-velocity component.


\subsection{Water Maser Distributions}

\begin{figure*}[h]
\begin{center}
\FigureFile(160mm,160mm){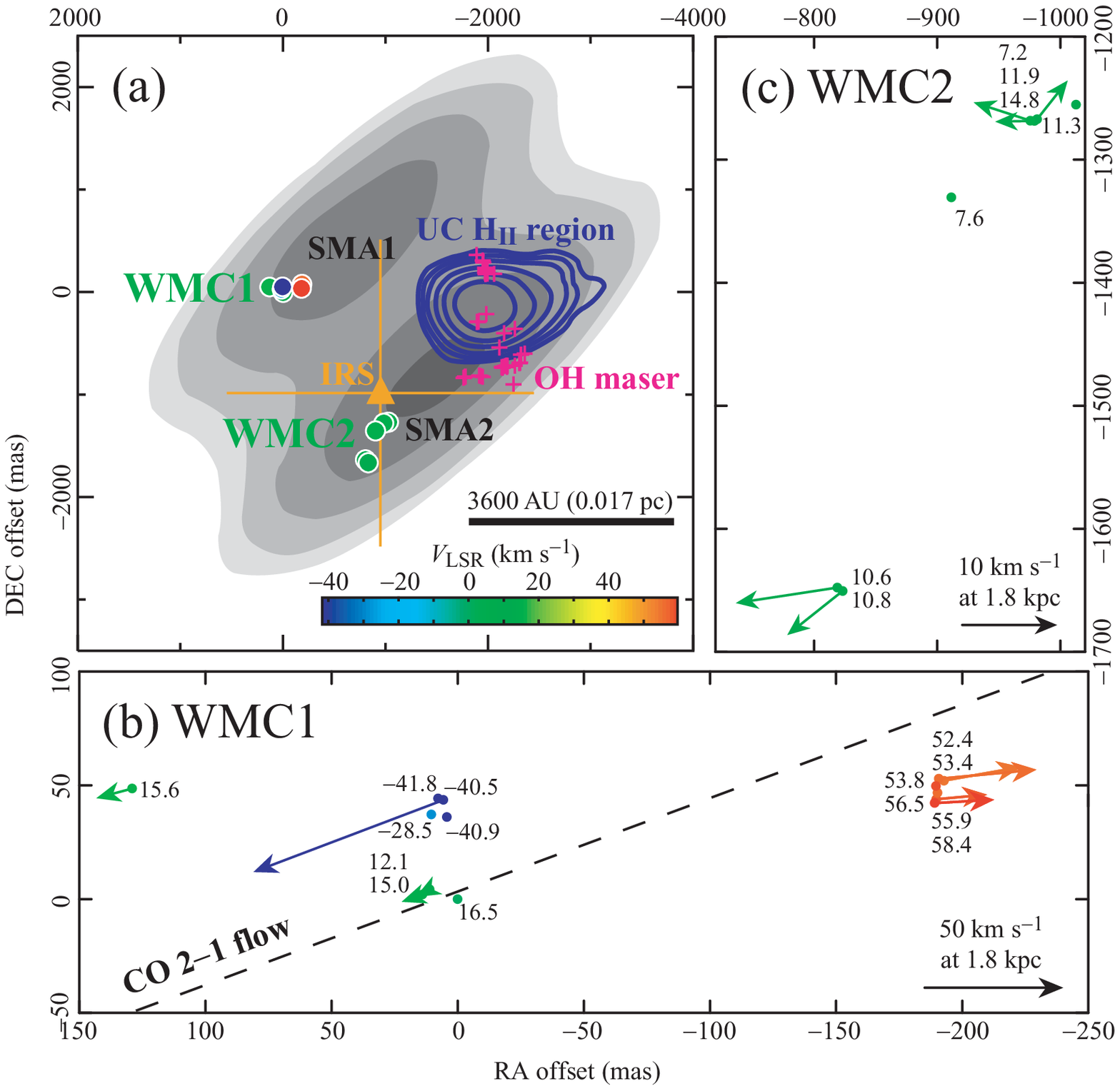}
\end{center}
\caption{(a):
         Observed water maser distributions (colored filled circle)
         superimposed on
         an 8.4 GHz radio continuum map (thick contour)
         of the UC H\emissiontype{II} region (\cite{argon}).
         The grey-scale map shows the 345 GHz radio continuum showing dust emission (\cite{su}),
         where the lowest grey contour indicates a half of the peak brightness.
         Crosses show OH maser distribution (\cite{fish}) and
         triangle indicates the 10$\mu$m infrared source (\cite{kumar2003}).
         The map origin with RA and DEC offsets = (0, 0) is at
         $\alpha$(J2000)=\timeform{20h10m09.201s} $\pm$ \timeform{0.004s} and
         $\delta$(J2000)=\timeform{31D31'36.02"} $\pm$ \timeform{0.08"}.
         1000 mas corresponds to 1800 AU at a distance of 1.8 kpc.
         (b), (c): Close-up to the two water maser clusters (WMC1 and WMC2) with proper motion vectors.
         The colored arrows and number added for each feature denote LSR velocity.
         The arrows at the bottom-rihgt corner in (b) and (c)
         indicate a proper motion of
         5.9 mas yr$^{-1}$ and 1.2 mas yr$^{-1}$ (50 km s$^{-1}$ and 10 km s$^{-1}$), respectively.
	 The dashed line in (b) indicates
         the axis of bipolar outflow observed in
         $^{12}$CO $J$ = 2--1 line (\cite{kumar2004})}
\label{fig:2}
\end{figure*}

Figure \ref{fig:2} shows distributions and the
proper motion vectors of water masers in ON1 (for proper motions, see next subsection).
21 features detected at the first epoch are plotted.
The color index denotes the LSR velocity range
from $-41.8$ to 58.4 km s$^{-1}$,
where 21 features are located.
The map origin is located at the position
of the reference maser feature at $V_{\rm LSR} = 16.5$ km s$^{-1}$,
which is estimated to be
$\alpha$(J2000)=\timeform{20h10m09.201s} $\pm$ \timeform{0.004s} and
$\delta$(J2000)=\timeform{31D31'36.02"} $\pm$ \timeform{0.08"}
from fringe rate analysis.
ON1 has two clusters of water maser features 
located at $(X,~Y) \simeq (\timeform{0"},~\timeform{0"})$
(hereafter WMC1 = water maser cluster 1)
and at $(X,~Y) \simeq (\timeform{-0".9},~\timeform{-1".4})$ (hereafterWMC2).
The separation of WMC1 and WMC2 is \timeform{1".6},
which corresponds to 2900 AU.

WMC1 is 
distributed within a region of
$\sim 320 \times 50$ mas ($580 \times 90$ AU).
The blue- ($V_{\rm LSR} = -41.8$ to $-28.5$ km s$^{-1}$) 
and red-shifted ($V_{\rm LSR} =$ 52.4--58.4 km s$^{-1}$) 
maser features appeared in WMC1.
Their separation is 195 mas, which corresponds to 350 AU.
The blue- and red-shifted maser features
are distributed within
areas of 9 $\times$ 11 mas (16 $\times$ 20 AU) and
5 $\times$ 14 mas (9 $\times$ 25 AU), respectively,
and were not detected outside these regions.
The systemic velocity of WMC1 is
$V_{\rm LSR} = 9.3 \pm 6.8$ km s$^{-1}$,
derived from 
three mean velocities of 
blue-shifted ($V_{\rm LSR} = -40.5 \pm 4.3$ km s$^{-1}$), 
red-shifted ($V_{\rm LSR} = 55.4 \pm 2.0$ km s$^{-1}$), and
low-velocity ($V_{\rm LSR} = 13.1 \pm 3.2$ km s$^{-1}$) features
detected in three epoch observations.

WMC2 is 
distributed within a region of 
$\sim 160 \times 380$ mas ($290 \times 680$ AU).
Only the low-velocity features 
($V_{\rm LSR} =$ 7.2--14.8 km s$^{-1}$) are detected in WMC2.
The systemic velocity of WMC2 is derived to be
$V_{\rm LSR} = 11.3 \pm 2.3$ km s$^{-1}$
from the mean velocity of low-velocity features 
detected in three epoch observations.

\subsection{Proper Motions}

\begin{figure*}[h]
\begin{center}
\FigureFile(80mm,80mm){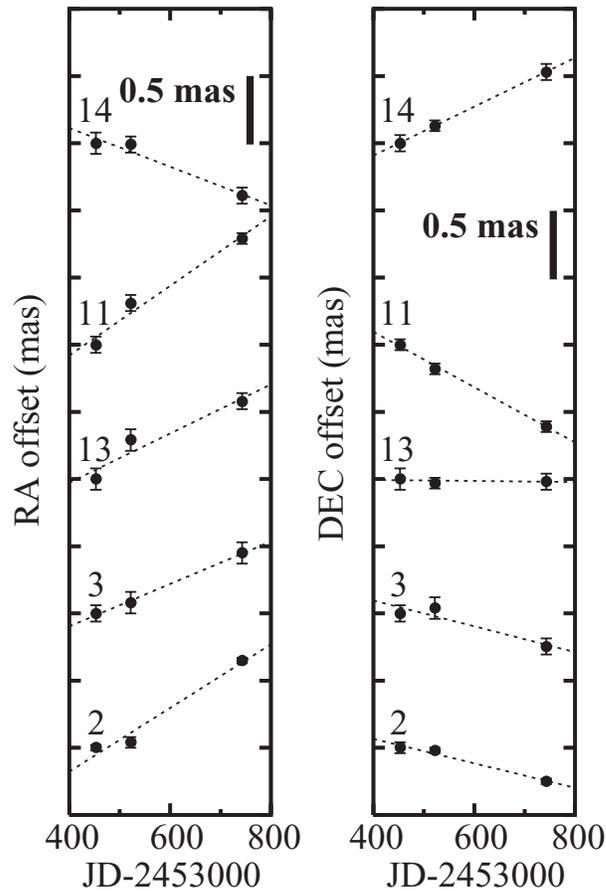}
\end{center}
\caption{Observed relative proper motions of water maser features in ON1.
        The proper motions of only the maser features detected at all
        three epochs are presented.
        A number added for each sub-panel shows 
        the assigned one listed in Table \ref{tab:2}.
        The dash line indicates a least-squares-fitted line assuming a
        constant velocity motion.}
\label{fig:3}
\end{figure*}

Table \ref{tab:2} lists 
observed proper motions of 14 maser features in ON1.
The maser features in different epochs were identified as 
the {\it same} feature,
if their LSR velocities were equal to 
each other within 0.42 km s$^{-1}$ (2-channel) and
if their positions were coincident 
within 2.5 mas at the first to second epoch 
and 7 mas at the second to third epoch.
The spatial ranges of 2.5 and 7 mas correspond to 
the proper motion of 100 km s$^{-1}$ (12 mas yr$^{-1}$).
Based on these identification criteria, 
each maser feature was identified in at least two epochs.
The proper motions have been calculated by performing a
linear least-squares fit of the positional offsets to the elapsed time.
Figure \ref{fig:3} shows observed time variations 
of right ascension and declination 
offsets (relative to feature ``5'')
of five features detected at all three epochs.

The proper motions of WMC1 
exhibit a bipolar outflow structure in the east-west direction. 
The proper motions show high ($69 \pm 11$ km s$^{-1}$) and 
low ($\sim$ 10 km s$^{-1}$)
expansion velocities.
The blue- and red-shifted features represent
the high-velocity and collimated outflow.
For the blueshifted features,
the proper motion of the brightest feature at
$V_{\rm LSR} = -40.5$ km s$^{-1}$ was obtained to be
$(\mu_x,~\mu_y)=$(7.9, $-$3.3) mas yr$^{-1}$,
which corresponds to
$(V_x,~V_y)=(68,~-28)$ km s$^{-1}$.
The mean proper motion of four redshifted features 
($V_{\rm LSR} = $ 52.4--58.4 km s$^{-1}$) is
$(\mu_x,~\mu_y)=(-3.0 \pm 0.7,~0.4 \pm 0.2)$ mas yr$^{-1}$,
which corresponds to
$(V_x,~V_y)=(-25 \pm 6,~3 \pm 2)$ km s$^{-1}$.
These proper motions  
appear to be associated 
with a common origin.

The simplest explanation for this fact is that 
the water masers in WMC1 are associated
with a bipolar outflow 
which is ejected from a YSO located
at the midpoint of the blue- and red-shifted features.
The expansion velocity between the blue-
and red-shifted features was estimated to be
$69 \pm 11$ km s$^{-1}$, from
the differences of their proper motions 
$(\Delta V_x,~\Delta V_y)=(93 \pm 6,~31 \pm 2)$ km s$^{-1}$
and LSR velocities $\Delta V_{\rm LSR} = 95 \pm 3$ km s$^{-1}$.
Inclination angle of the direction of expansion 
was determined to be 44 $\pm$ \timeform{3D}.
The position angle and opening angle derived from the
distributions of blue- and red-shifted features are
\timeform{92D} and \timeform{10D}, respectively.
The LSR velocities of three low-velocity features
($V_{\rm LSR} =$ 12.1, 15.0, 15.6 km s$^{-1}$) 
in WMC1 show low expansion velocity ($\sim$ 10 km s$^{-1}$).
Their LSR velocities are
close to 
the LSR velocity of the quiescent molecular cloud 
observed in NH$_3$ line ($V_{\rm LSR} \sim$ 10 km s$^{-1}$; \cite{zheng}).
This may indicate an interaction between the outflow and
the dense surrounding cloud (\cite{gen81}).
Their proper motions are 
within 
$\mu_x =$ 0.5--1.5, $\mu_y = -0.5$ to $-0.3$ mas yr$^{-1}$,
which correspond to
$V_x =$ 4--13, $V_y = -4$ to $-3$ km s$^{-1}$.

The proper motions of WMC2 show $\mu \sim 1$ mas yr$^{-1}$
($V \sim 10$ km s$^{-1}$).
Althouth it is unknown what is associated with WMC2,
the proper motions of WMC2 are not originated to WMC1.
This indicates the presence of a driving source 
which is different from WMC1.
Therefore, our proper motion measurements represent that
there are at least two driving sources of water masers in ON1.


\section{Discussion}

\subsection{Driving Sources of Water Masers}

Figure \ref{fig:2}(a) shows positions of the water masers relative
to the 8.4 GHz continuum emission.
The water masers are located at \timeform{2"} east
from the peak of 8.4 GHz continuum emission
with an absolute position accuracy of $\approx$ \timeform{0".3} (\cite{argon}).
The water masers
are not coincident with the UC H\emissiontype{II} region
and appear to be associated with the 
YSO formed on the east side of the UC H\emissiontype{II} region.

Submillimeter continuum emission traces dust emission
around a YSO. 
The 345 GHz continuum emission observed with SMA
has two components of north-eastern (SMA1) 
and south-western (SMA2) ones (\cite{su}).
The water masers would be associated with SMA1 and SMA2.
Water masers appear to have a position offset of approximately
\timeform{1"} south-east from the both peak positions of 
345 GHz continuum emission.
This offset would be due to an insufficient angular resolution
of the 345 GHz continuum observation of $\sim$ \timeform{3"}
and insufficient absolute position accuracy.

A 10.5 $\mu$m infrared source, which is not detected in
the 2.2 $\mu$m and 3.75 $\mu$m emissions (\cite{kumar2003}), is
situated near the center of
the two WMCs and UC H\emissiontype{II} region.
Because the 10.5 $\mu$m emission is extended
in a \timeform{3"} $\times$ \timeform{3"} area,
it is unclear whether 
this is associated with the present water maser features.

The total far-infrared luminosity of ON1
derived from the IRAS data is
$10^{4.1} \LO$ and
the luminosity of UC H\emissiontype{II} region
observed in 1.3 to 20 cm radio continuum emission
is $10^{4.2} \LO$ (\cite{macleod}).
The agreement of the luminosities of far-infrared and
radio continuum emissions
indicates that ON1 has a single luminous star of spectral type of B0
($L \geq 10^{4} \LO$)
which forms the UC H\emissiontype{II} region.
The luminosities of YSOs in WMC1 and WMC2 seem to be lower 
($< 10^4 \LO$) 
than that of the B0 star exciting the UC H\emissiontype{II} region.

We may thus conclude that
the water masers WMC1 and WMC2 are associated with
two YSOs on the eastern side of UC H\emissiontype{II} region.
These two YSOs are 
still surrounded by dusty envelopes,
emitting the 345 GHz submillimeter emission.


\subsection{Outflow and YSO in WMC1}

The outflow of WMC1 is coincident with
a jet-like outflow 
extended 
in the east-west direction 
by $\sim$ 0.07 pc,
which was found by IRAM observations in the $^{12}$CO $J =$ 2--1 line (\cite{kumar2004}).
The dynamical age of CO 2--1 outflow is
estimated to be (5--7) $\times 10^{3}$ yr
using the LSR velocity difference from the systemic velocity of CO 2--1
(12 km s$^{-1}$)
and the inclination angle of the outflow obtained by the present work ($44 \pm \timeform{3D}$).
The velocity spans in
$^{12}$CO $J$ = 1--0 and 2--1 lines are
35 km s$^{-1}$ and 24 km s$^{-1}$, respectively
(\cite{xu06}; \cite{kumar2004}).
These values are smaller than the velocity span of water maser ($\sim$ 100 km s$^{-1}$).
This may be 
because the size of the high
velocity outflow observed in the water masers ($< \timeform{0."2}$)
is too compact to be
detected with the beam of the CO 1--0 ($\sim$ \timeform{15"})
and CO 2--1 ($\sim$ \timeform{2"}) lines.
The dynamical age of the CO outflow of (5--7) $\times 10^3$ yr
suggests that the age of the YSO in WMC1 is about $\sim 10^4$ yr.

Water maser features are most likely located in
shock regions where the outflow from YSO hits ambient gases.
We derive the momentum rate of the outflow in WMC1
using the method of \citet{tor03}.
The momentum rate is given by,
\begin{eqnarray}
\dot{P}_{\rm f} = 2 \Omega_{\rm f} R^2 \rho_{\rm f} V_{\rm f}^2,
\end{eqnarray}
where
$\Omega_{\rm f} = 2 \pi (1-\cos(\theta_{\rm op}/2))$ is the solid angle of the outflow,
$R$ the distance from the star,
$\rho_{\rm f}$ the mass density in the outflow,
$V_{\rm f}$ the expansion velocity of the outflow.
Assuming a typical gas density necessary for
water masering,
$n$(H$_2$) = $10^8$ cm$^{-3}$ (\cite{eli92}),
the momentum rate of outflow can be estimated to be
(1--2) $\times 10^{-3} \MO$ yr$^{-1}$ km s$^{-1}$
for the observed expansion velocity ($V_{\rm f} = 69 \pm 11$ km s$^{-1}$),
the distance from the star ($R$ = 175 AU/cos(\timeform{44D}) = 245 AU),
and the opening angle ($\theta_{\rm op} = \timeform{10D}$).
In addition, using this momentum rate
and dynamical age derived from the CO 2--1 outflow, 
we can estimate the momentum of the outflow in WMC1 to be
5--14 $\MO$ km s$^{-1}$.
This value is 10--30\% of the momentum 
of the largest outflow 
in ON1 with a size of 0.94 pc
found in $^{12}$CO $J$ = 1--0 line (\cite{xu06}).
Therefore, the outflow in WMC1 would not mainly contribute to
the formation of the CO 1--0 outflow.


\subsection{Star Formation Activity of ON1}

Dense molecular gas, which is traced in
CS $J$ = 5--4 line (\cite{shi03})
and dust emission at 350 $\mu$m (\cite{mue02}),
extend to the 
east and north side of the UC H\emissiontype{II} region.
The different locations of the UC H\emissiontype{II} region,
submillimeter continuum emissions, water masers, and dense molecular cloud
suggest that the star formation activity of ON1
moves from west to east.
In Table \ref{tab:3}, we summarize the possible properties
of B0 star exciting the UC H\emissiontype{II} region 
as well as the YSO associated with WMC1 and WMC2.
The mass of star in UC H\emissiontype{II} region is $\sim 15 \MO$,
which indicated by the spectral type of apploximately B0 (\cite{macleod}).
The mass of YSO in WMC1 would be $\sim$ 2--15 \MO, because
the momentum rate of the outflow in WMC1 corresponds to a typical value 
of intermediate-mass YSO (\cite{she05}).
The difference between WMC1 and WMC2
is only the expansion velocity of water masers.
This velocity difference may reflect the difference
of outflow energy between WMC1 and WMC2.
This indicates lower power,
implying a lower mass of the forming YSO in WMC2 than
in WMC1.

Finally, we propose that the B0 star is the possible
driving source of the CO 1--0 outflow.
The dynamical age of CO 1--0 outflow 
is estimated to be $7.3 \times 10^{4}$ yr
at our assumed distance (\cite{xu06}).
Thus, we think the age of the B0 star 
is $\sim 10^5$ yr.

\subsection{A Binary System Formed by WMC1 and UC H\emissiontype{II} Region?}

The LSR velocities of WMC1 and UC H\emissiontype{II} region
is estimated to be $V_{\rm LSR} = 9.3 \pm 6.8$, and
$12.3 \pm 3.5$ km s$^{-1}$, from the systemic velocities of
water masers and OH masers (\cite{fish}), respectively.
If WMC1
and UC H\emissiontype{II} region were separated at a relative velocity of 3 km s$^{-1}$ 
during the formation timescale of B0 star exciting the UC H\emissiontype{II} region of $\sim 10^5$ yr,
WMC1 and UC H\emissiontype{II} region should already be separaed by 
$\sim$ 60000 AU (0.30 pc).  
However, their separation on the sky is 3600 AU (0.017 pc).
Therefore,
we may consider that
WMC1 and UC H\emissiontype{II} region are gravitationally bound.
Assuming that WMC1 and 
UC H\emissiontype{II} region
are gravitationally bound,
that total mass ($M_{\rm t}$) can be estimated by,
\begin{eqnarray}
M_{\rm t} & \sim & \frac{R v^{2}}{G} \\
          & \sim & 4.1 \times \left( \frac{R}{3600~{\rm AU}} \right) 
            \left( \frac{v}{1~{\rm km~s}^{-1}} \right)^2 \MO 
\end{eqnarray}
where, $R$ is a separation of the WMC1 and UC H\emissiontype{II} region
and $v$ is the relative velocity.
This total mass is minimum value,
because the separation along the line of sight and
the relative velocity of the sky are still unknown.
The total mass is derived $M_{\rm t} \sim 37 \MO$
from the separation of 3600 AU and the relative velocity of $v = 3$ km s$^{-1}$.
The masses of 345 GHz dust emission core associated with 
the WMC1 and UC H\emissiontype{II} region
are estimated to be
6 \MO, and 6 \MO, respectively (\cite{su06}). 
Therefore, the total mass YSO and B0 star, and accompanying dust-emission cores
is $\sim$ 29--42 \MO.
This value is consistent with the total mass
estimated on the basis of assumption 
of a gravitationally bound system.
Therefore,
we propose that the YSO in WMC1 and B0 star exciting the UC H\emissiontype{II} region
form a binary star. 
The LSR velocity of WMC2 ($V_{\rm LSR} = 11.3 \pm 2.3$ km s$^{-1}$)
indicates that
the YSO in WMC2 is a third object of this bound system.

\begin{figure*}[h]
\begin{center}
\FigureFile(100mm,80mm){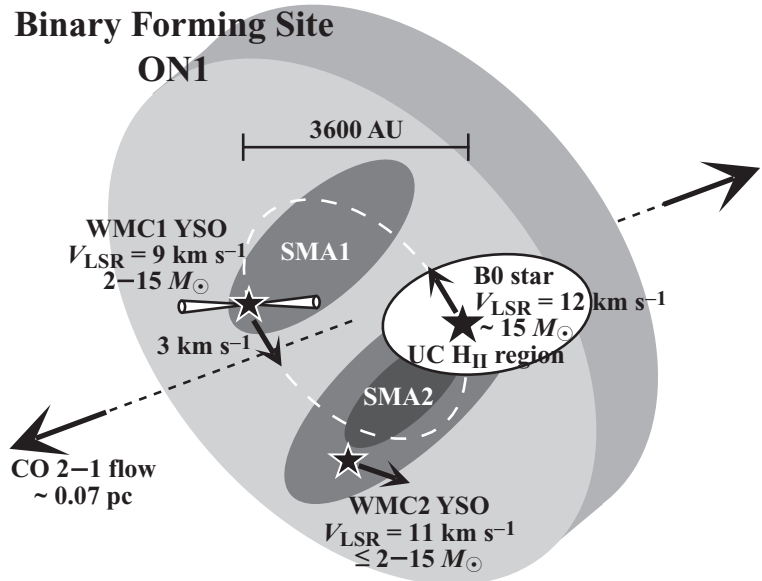}
\end{center}
\caption{Schematic representation of the structure of ON1.}
\label{fig:4}
\end{figure*}

In Figure \ref{fig:4}, we illustrate a possible structure
of ON1, as
inferred from the present consideration based on Figure \ref{fig:2}(a).
A rotating disk is found in NH$_3$ $(J,K) = (1,1)$, 
and H$^{13}$CO$^{+} J =$ 1--0 lines
with the VLA and the BIMA array (\cite{zheng}; \cite{lim02}).
The orbit in Figure \ref{fig:4} is assumed to be at a
position angle of \timeform{40D} which is 
seen in the velocity gradient in H$^{13}$CO$^{+}$ line (\cite{lim02}).


\section{Conclusions}

The following conclusions are drawn from this study:

\begin{enumerate}

\item
We carried out three epoch observations of the ON1
water masers with the JVN, and successfully
measured the proper motions of ON1 water masers
for the first time.

\item
ON1 has two major water maser clusters (WMC1 and WMC2) 
which are separated by 2900 AU.
Both the WMCs are located at 3600 AU from 
the UC H\emissiontype{II} region seen at 8.4 GHz continuum emission.

\item
Proper motion measurement reveals that
WMC1 is associated with a bipolar outflow 
elongated in the east-west direction
with high expansion velocity of $69 \pm 11$ km s$^{-1}$.
This outflow is coincident with
a jet-like outflow found in CO 2--1 line.

\item
The WMC1 and WMC2 are associated 
with two 345 GHz continuum sources
on the east side of the UC H\emissiontype{II} region. 
These two YSOs are 
still surrounded by dusty envelopes.

\item
The star formation activity of ON1 appears to move
from west side of the UC H\emissiontype{II} region to 
east side of two WMCs and dense molecular cloud 
observed in CS lines.

\item
We suggests that the WMC1 and UC H\emissiontype{II} region in ON1
form a binary system.
The relative velocity and 
total mass of WMC1 and UC H\emissiontype{II} region
are estimated to be $\Delta V_{\rm LSR} \sim 3$ km s$^{-1}$ 
and $M_{\rm t} \sim 37 \MO$, respectively.

\end{enumerate}


\bigskip

The authors wish to thank to all staff members and students of 
VERA team for observing assistance and support.
We hope VERA observe one thousand of water maser sources to perform the 
astrometry of galactic star forming regions and 
investigate the structure and dynamics of the Milky Way Galaxy.
We thank an anonymous referee for very useful comments.
T.O. and H.I. were supported by a Grant-in-Aid for Scientific Research 
from Japan Society for Promotion Science (17340055).


\begin{table*}[h]
\begin{center}
\caption{Status of the telescopes, data reduction, and resulting
performances in the individual epochs of the JVN observations.} 
\label{tab:1}
\begin{tabular}{cccccccc}
\hline \hline
\multicolumn{1}{c}{Epoch} & 
\multicolumn{1}{c}{Date} & 
\multicolumn{1}{c}{Duration} & 
\multicolumn{1}{c}{Used} & 
\multicolumn{1}{c}{Reference} & 
\multicolumn{1}{c}{1-$\sigma$ level} & 
\multicolumn{1}{c}{Synthesized} & 
\multicolumn{1}{c}{Number of} \\ 
 &
 &
 &
\multicolumn{1}{c}{telescopes\footnotemark[$*$]} & 
\multicolumn{1}{c}{velocity\footnotemark[$\dagger$]} & 
\multicolumn{1}{c}{noise} & 
\multicolumn{1}{c}{beam\footnotemark[$\ddagger$]} & 
\multicolumn{1}{c}{detected} \\ 
 &
 &
\multicolumn{1}{c}{(hr)} & 
 &
\multicolumn{1}{c}{(km s$^{-1}$)} & 
\multicolumn{1}{c}{(Jy beam$^{-1}$)} & 
\multicolumn{1}{c}{(mas)} & 
\multicolumn{1}{c}{features} \\ 
\hline
1 ... & 2005 Mar 24 & 10 & MZ, IR, OG, IS, KS\footnotemark[$\S$], NB & 16.5 & 0.045 & $1.9 \times 0.7,~\timeform{-38D}$ & 21 \\ 
2 ... & 2005 Jun 1  & 10 & MZ, OG, IS, KS, NB     & 16.5 & 0.032 & $3.4 \times 1.3,~\timeform{-86D}$ & 18 \\ 
3 ... & 2006 Jan 8  & 10 & MZ, IR, OG, IS, KS     & 16.5 & 0.039 & $1.7 \times 0.9,~\timeform{-51D}$ & 23 \\ 
\hline
\multicolumn{8}{@{}l@{}} {\hbox to 0pt{\parbox{170mm} {\footnotesize
\footnotemark[$*$] 
	Telescopes that were effectively operated and whose recorded data were valid:
        MZ: the VERA 20-m telescope at Mizusawa,
        IR: the VERA 20-m telescope at Iriki,
        OG: the VERA 20-m telescope at Ogasawara Is,
        IS: the VERA 20-m telescope at Ishigakijima Is,
        KS: the NiCT 34-m telescope at Kashima,
        NB: the NRO 45-m telescope at Nobeyama.
\par\noindent
\footnotemark[$\dagger$]
	Local-standard-of-rest velocity of the spectral 
	channel used for the phase reference in data reduction.
\par\noindent
\footnotemark[$\ddagger$]
	The synthesized beam made in natural weight; 
        major and minor axis lengths and position angle.
\par\noindent
\footnotemark[$\S$]
	Ceasing operation for 8 hours due to strong winds.
\par\noindent
}\hss}}
\end{tabular}
\end{center}
\end{table*}

\begin{table*}[h]
\begin{center}
\caption{Parameters of the water maser features 
         identified by proper motion toward ON1.}
\label{tab:2}
\begin{tabular}{rrrrrrrrrrrrr}
\hline \hline
\multicolumn{1}{c}{ID\footnotemark[$*$]} & 
\multicolumn{4}{c}{Offset\footnotemark[$\dagger$]} & 
\multicolumn{1}{c}{LSR velocity} & 
\multicolumn{4}{c}{Proper motion\footnotemark[$\dagger$]} & 
\multicolumn{3}{c}{Peak intensity at three epochs} \\ 
   & 
\multicolumn{4}{c}{(mas)}           & 
\multicolumn{1}{c}{(km s$^{-1}$)}   & 
\multicolumn{4}{c}{(mas yr$^{-1}$)} & 
\multicolumn{3}{c}{(Jy beam$^{-1}$)} \\
   &
\multicolumn{4}{c}{\hrulefill \ }   & 
\multicolumn{1}{c}{\hrulefill \ }   & 
\multicolumn{4}{c}{\hrulefill \ }   & 
\multicolumn{3}{c}{\hrulefill \ }   \\ 
   &
\multicolumn{1}{c}{$X$} &
\multicolumn{1}{c}{$\sigma X$} &
\multicolumn{1}{c}{$Y$} &
\multicolumn{1}{c}{$\sigma Y$} &
\multicolumn{1}{c}{$V_{\rm LSR}$} &
\multicolumn{1}{c}{$\mu_{x}$} &
\multicolumn{1}{c}{$\sigma \mu_{x}$} &
\multicolumn{1}{c}{$\mu_{y}$} &
\multicolumn{1}{c}{$\sigma \mu_{y}$} &
\multicolumn{1}{c}{Epoch 1} &
\multicolumn{1}{c}{Epoch 2} &
\multicolumn{1}{c}{Epoch 3} \\
\hline
1  &    129.04 & 0.05 &      48.58 & 0.05 &    15.69 &    1.49 & ...  & $-$0.44 & ...  &  0.40 &  ...  &  0.65 \\
2  &     13.99 & 0.01 &       1.98 & 0.02 &    15.06 &    0.86 & 0.15 & $-$0.33 & 0.05 & 13.70 & 15.00 & 21.50 \\
3  &     11.04 & 0.03 &       4.29 & 0.03 &    12.11 &    0.57 & 0.03 & $-$0.34 & 0.13 &  1.03 &  0.81 &  0.56 \\
4  &      5.69 & 0.06 &      43.55 & 0.08 & $-$40.54 &    7.93 & ...  & $-$3.33 & ...  &  0.53 &  6.82 &  ...  \\
5  &      0.00 & 0.06 &       0.00 & 0.04 &    16.53 &    0.00 & ...  &    0.00 & ...  &  6.55 &  6.75 &  5.28 \\
6  & $-$189.05 & 0.02 &      42.17 & 0.02 &    58.45 & $-$2.48 & ...  &    0.15 & ...  &  3.09 &  5.50 &  ...  \\
7  & $-$189.71 & 0.02 &      43.91 & 0.02 &    55.93 & $-$2.11 & ...  &    0.21 & ...  &  7.53 &  3.69 &  ...  \\
8  & $-$190.73 & 0.01 &      52.90 & 0.01 &    53.40 & $-$4.07 & ...  &    0.42 & ...  & 10.60 & 17.90 &  ...  \\
9  & $-$192.73 & 0.01 &      52.05 & 0.01 &    52.56 & $-$3.22 & ...  &    0.63 & ...  & 28.00 & 26.00 &  ...  \\
10 & $-$818.83 & 0.13 & $-$1647.78 & 0.07 &    10.64 &    1.73 & ...  & $-$0.26 & ...  &  ...  &  1.63 &  1.90 \\
11 & $-$823.19 & 0.03 & $-$1650.52 & 0.02 &    10.85 &    0.94 & 0.16 & $-$0.75 & 0.04 &  2.96 &  2.67 &  3.69 \\
12 & $-$975.58 & 0.07 & $-$1268.37 & 0.06 &    14.85 &    0.96 & ...  &    0.31 & ...  &  1.31 &  ...  &  0.66 \\
13 & $-$978.89 & 0.04 & $-$1268.50 & 0.04 &    11.90 &    0.66 & 0.20 & $-$0.01 & 0.03 &  0.89 &  1.16 &  0.71 \\
14 & $-$981.02 & 0.04 & $-$1267.26 & 0.03 &     7.27 & $-$0.52 & 0.11 &    0.66 & 0.00 &  2.23 &  1.30 &  0.86 \\
\hline
\multicolumn{13}{@{}l@{}} {\hbox to 0pt{\parbox{170mm}{\footnotesize
\footnotemark[$*$]
        Feature ID number.
\par\noindent
\footnotemark[$\dagger$]
        Relative value with respect to the position-reference maser feature: ID 5.
\par\noindent
}\hss}}
\end{tabular}
\end{center}
\end{table*}

\begin{table*}[h]
\begin{center}
\caption{Properties of B0 star in UC H\emissiontype{II} region
         and YSOs in WMC1 and WMC2.}
\label{tab:3}
\begin{tabular}{cccccccc}
\hline \hline
\multicolumn{1}{c}{No} &
\multicolumn{1}{c}{Name} &
\multicolumn{1}{c}{Emission} &
\multicolumn{1}{c}{Maser} &
\multicolumn{1}{c}{Velocity span\footnotemark[$*$]} &
\multicolumn{1}{c}{Luminosity} &
\multicolumn{1}{c}{Age\footnotemark[$\dagger$]} &
\multicolumn{1}{c}{Mass\footnotemark[$\ddagger$]} \\
 &
 &
 &
 &
\multicolumn{1}{c}{(km s$^{-1}$)} &
\multicolumn{1}{c}{(\LO)} &
\multicolumn{1}{c}{(yr)}  &
\multicolumn{1}{c}{(\MO)} \\
\hline
1 ...      & 
B0 star    & 
Ionized gas\footnotemark[$\S$] + Dust\footnotemark[$\|$] &
OH\footnotemark[$\#$] &
$\sim 15$\footnotemark[$\#$] &
$10^{4.2}$\footnotemark[$**$] &
$\sim 10^5$ &
$\sim 15$ \\
2 ...      &
WMC1 YSO   & 
Dust\footnotemark[$\S$] &
Water      &
$\sim 100$ &
$< 10^4$\footnotemark[$\dagger\dagger$] &
$\sim 10^4$ &
$\sim$ 2--15 \MO \\
3 ...      &
WMC2 YSO   &
Dust\footnotemark[$\|$] &
Water      &
$\sim 10$  &
$< 10^4$\footnotemark[$\dagger\dagger$] &
...        &
$\leq$ 2--15 \MO \\
\hline
\multicolumn{7}{@{}l@{}} {\hbox to 0pt{\parbox{170mm}{\footnotesize
\footnotemark[$*$]
        LSR velocity span of masers.
\par\noindent
\footnotemark[$\dagger$]
        Estimated ages of B0 star and WMC1 YSO (see subsection 4.2 and 4.3).
\par\noindent
\footnotemark[$\ddagger$]
        Estimated masses (see subsection 4.3).
\par\noindent
\footnotemark[$\S$]
        Traced by 8.4 GHz continuum emission (\cite{argon}).
\par\noindent
\footnotemark[$\|$]
        Traced by 345 GHz continuum emission (\cite{su}).
\par\noindent
\footnotemark[$\#$]
	Results obtaind in OH maser observation (\cite{fish}).
\par\noindent
\footnotemark[$**$]
	Luminosity obtained in 1.3 to 20 cm radio continuum emissions (\cite{macleod}).
\par\noindent
\footnotemark[$\dagger\dagger$]
	Luminosity estimated from the far-infrared and radio continuum emissions (see subsection 4.1).
\par\noindent
}\hss}}
\end{tabular}
\end{center}
\end{table*}

\begin{thebibliography}{}
\bibitem[Andr\'e et al.(1993)]{and93}
   Andr\'e, P., Ward-Thompson, D., \& Barsony, M.
   1993, ApJ, 406, 122
\bibitem[Argon et al.(2000)]{argon}
   Argon, A. L., Reid, M. J., \& Menten, K. M. 
   2000, ApJS, 129, 159
\bibitem[Cesaroni et al.(1988)]{cesaroni}
   Cesaroni, R., Palagi, F., Felli, M., Catarzi, M., Comoretto, G.,
   di Francos, Giovanardi, C., \& Palla, F.
   1988, A\&AS, 76, 445
\bibitem[Chikada et al.(1991)]{chikada}
   Chikada, Y., et al. 
   1991, in Frontiers of VLBI, ed.
   H. Hirabayashi, M. Inoue, \& H. Kobayashi 
   (Tokyo: Uiversal Academy Press), 79
\bibitem[Doi et al.(2006)]{doi}
   Doi, A., et al. 
   2006, PASJ, 58, 777
\bibitem[Downes et al.(1979)]{downes}
   Downes, D., Genzel, R., Moran, J. M., Johnston, K. J., Matveenko, L. I.,
   Kogan, L. R., Kostenko, V. I., \& Ronnang, B. 
   1979, A\&A, 79, 233
\bibitem[Elizur et al.(1992)]{eli92}
   Elitzur, M., Hollenbach, D. J., \& McKee, C. F.
   1992, ApJ, 394, 221
\bibitem[Fish et al.(2005)]{fish}
   Fish, V. L., Reid, M. J., Argon, A. L., \& Zheng, X. W.
   2005, ApJS, 160, 220
\bibitem[Genzel \& Downes(1977)]{genzel77}
   Genzel, R., \& Downes, D.
   1977, A\&AS, 30, 145
\bibitem[Genzel et al.(1981)]{gen81}
   Genzel, R., Reid, M. J., Moran, J. M., \& Downes, D. 
   1981, ApJ, 244, 884
\bibitem[Imai et al.(2000)]{imai}
   Imai, H., Kameya, O., Sasao, T., Miyoshi, M., 
   Deguchi, S., Horiuchi, S., \& Asaki, Y.
   2000, ApJ, 538, 751
\bibitem[Imai et al.(2006)]{imai06}
   Imai, H., Omodaka, T., Hirota, T., Umemoto, T., Sorai, K., \& Kondo, T.
   2006, PASJ, 58, 883 
\bibitem[Kumar et al.(2002)]{kumar2002}
   Kumar, M. S. N., Bachiller, R., \& Davis, C. J. 
   2002, ApJ, 576, 313
\bibitem[Kumar et al.(2003)]{kumar2003}
   Kumar, M. S. N., Davis, C. J., \& Bachiller, R.
   2003, Ap\&SS, 287, 191
\bibitem[Kumar et al.(2004)]{kumar2004}
   Kumar, M. S. N., Tafalla M., \& Bachiller, R. 
   2004, A\&A, 426, 195
\bibitem[Kurtz \& Hofner(2005)]{kurtz}
   Kurtz, S., \& Hofner, P.
   2005, AJ, 130, 711
\bibitem[Lim et al.(2002)]{lim02}
   Lim, J., Choi, M., \& Ho, P. T. P.
   2002, ASPC, 267, 385
\bibitem[MacLeod et al.(1998)]{macleod}
   MacLeod, G. C., Scalise, E. J., Saedt, S., Galt, J. A., \& Gaylard, M. J.
   1998, AJ, 116, 1897
\bibitem[Moscadelli et al.(2007)]{mos07}
   Moscadelli, L., Goddi, C., Cesaroni, R., Beltr\'an, M. T., \& Furuya, R. S.
   2007, A\&A, 472, 867
\bibitem[Mueller et al.(2002)]{mue02}
   Mueller, K. E., Shirley, Y. L., Evans, N. J., \& Jacobson, H R.
   2002, ApJS, 143, 469
\bibitem[Shepherd(2005)]{she05}
   Shepherd, D.
   2005, in IAU Symp. 227, ed. R. Cesaroni, M. Felli, E.
   Churchwelll, \& M. Walmsley
   (Cambridge: Cambridge Univ. Press), 237
\bibitem[Shirley et al.(2003)]{shi03}
   Shirley, Y. L., Evans, N. J. II, Young, K. E., Knez, C., \& Jaffe, D. T.
   2003, ApJS, 149, 375
\bibitem[Su et al.(2004)]{su}
   Su, Y. N., et al. 
   2004, ApJ, 616, 39
\bibitem[Su et al.(2006)]{su06}
   Su, Y. N., Liu, H. R., \& Lim, J.
   2006, 
   in the proceedings of the East Asian Young Astronomers Meeting 2006,
   Y. Urata, D. Kinoshita, T. Sekiguchi, A. Yonehara eds.
   (National Astronomical Observatory of Japan: Mitaka), p17
\bibitem[Torrelles et al.(2001)]{tor01}
   Torrelles, J. M., et al.
   2001, ApJ, 560, 853
\bibitem[Torrelles et al.(2003)]{tor03}
   Torrelles, J. M., et al.
   2003, ApJ, 598, L115
\bibitem[Xu et al.(2006)]{xu06}
   Xu, Y. et al.
   2006, AJ, 132, 20
\bibitem[Zheng et al.(1985)]{zheng}
   Zheng, X. W., Ho, P. T. P., Reid, M. J., \& Schneps, M. H. 
   1985, ApJ, 293, 522
\end{thebibliography}
\end{document}